\documentclass[aps,pra,twocolumn,amsmath,amssymb,superscriptaddress]{revtex4-2}
\usepackage[english]{babel}
\usepackage{latexsym}
\usepackage{graphics}
\usepackage{epsfig}
\usepackage{color}
\usepackage{bm}
\usepackage{amsmath}
\usepackage{amssymb}
\usepackage{physics}
\usepackage{hyperref}
\hypersetup{
   colorlinks = true,
   linkcolor=blue,
   citecolor=blue,
   urlcolor=blue
 }

\newcommand{\1}{\uparrow}
\newcommand{\2}{\downarrow}

\begin{document}

\title{Confinement-Induced Resonances in Rabi-Coupled Bosonic Mixtures}

\author{A. Tononi}
\affiliation{Department de F\'isica, Universitat Polit\`ecnica de Catalunya, Campus Nord B4-B5, E-08034, Barcelona, Spain}

\author{P. Massignan}
\affiliation{Department de F\'isica, Universitat Polit\`ecnica de Catalunya, Campus Nord B4-B5, E-08034, Barcelona, Spain}

\begin{abstract}
We consider coherently coupled bosonic mixtures scattering at low energies in the presence of an external confinement along either one or two directions.
We exactly solve the two-body scattering problem, showing that for large Rabi coupling the confinement-induced resonance can be displaced towards scattering lengths much smaller than the oscillator length. 
Our results make the observation of confinement-induced resonances more tunable and accessible, offering yet another handle for the efficient control of strong interactions in ultracold quantum gases.
\end{abstract}

\maketitle

\textbf{Introduction.---}
Low-energy scattering with short-range interparticle interactions is a fundamental problem of quantum physics.
It underlies our understanding of atomic interactions in the ultracold regime and enables the precise description of thermodynamics, few-body physics, and dynamics \cite{pethick2008}.
For dilute ultracold bosons, in particular, interactions are well described by zero-range models including only the partial $s$ wave \cite{pitaevskii2016}. 
The $s$-wave scattering problem of single-species gases has been solved in any spatial dimension $D$ \cite{Landau1991}, expressing the $3D$, $2D$, and $1D$ scattering amplitudes in terms of the $D$-dimensional scattering lengths.

Experimentally, low-dimensional systems are produced by confining the gas with strong harmonic potential along either one or two directions.
The scattering problem solution in such \textit{quasi}-low-dimensional setup displays confinement-induced resonances \cite{olshanii1998, petrovshlyapnikov, bergeman2003, massignan2006, pricoupenko2007, pricoupenko2007b, peng2010, oghittu2025}. 
Such a feature, absent in the purely $D$-dimensional scenario, is produced by virtual excitations of the scattering state to molecular trap states.
In particular, the resonance occurs when the $s$-wave scattering length $a$ is comparable to the transversal oscillator length $l_{\perp}$ and experimental observations were made possible by largely increasing the value of $a$ via Feshbach resonances \cite{haller2010}.

Interactions have also been characterized in detail for various two-component gases (for a general overview see \cite{baroni2024quantum}). Furthermore, scattering in ultracold atomic mixtures coupled by external fields has been analyzed in the context of 3D bosons with Rabi coupling \cite{lavoine2021, sanz2022}, of spin-orbit coupled fermions in quasi-1D \cite{zhang2013, zhang2014} and in quasi-2D \cite{wang2016}, of quasi-1D spin-orbit coupled mixtures with generic statistics \cite{wang2018}, {and in coupled multicomponent gases \cite{cui2014, bleu2025, zulli2025, mulkerin2024}}. 
While reaching the resonant regime in single-species gases may be difficult since generally $a \ll l_\perp$, external fields may facilitate the observation of confinement-induced resonances.

Here we exactly solve the scattering problem in 3D and in quasi-low-dimensional Rabi-coupled bosonic mixtures. 
We find that, by controlling the strength of the driving field, it is possible to displace the resonance location towards smaller 3D scattering lengths.
This phenomenon can be understood through a simple length scales comparison: the confinement-induced resonance, which occurs in the uncoupled case when one of the scattering lengths satisfies $a \sim l_{\perp}$, is shifted to smaller values of $a \sim l_{\Omega} = 1/\sqrt{\Omega} < l_{\perp}$ for sufficiently strong Rabi coupling $\Omega$.
This physically means that strong Rabi coupling facilitates the virtual excitation of colliding atoms to excited trap states, which is the fundamental mechanism underlying confinement-induced resonances.
In particular, for realistic experimental implementations \cite{lavoine2021}, the resonance location shifts from $a/l_{\perp} \sim 1$ to $a/l_{\perp} \sim 10^{-1}$, thus closer to the natural ratio achieved in trapped alkali-metal atoms.
These scattering properties can be analyzed in the available experiments with ultracold bosonic mixtures \cite{lavoine2021, chisholm2022}, for instance through measurements of atomic losses \cite{haller2010, capecchi2023} and spectroscopy \cite{moritz2005}, or studies of the condensate expansion \cite{eid2025}, by producing solitonic states \cite{frolian2022}, or by measuring collective modes of the condensate \cite{hammond2022}.
Clearly, observing such consequences requires careful control of magnetic-field fluctuations and Rabi-induced heating, which may reduce the condensate lifetime and can be mitigated, for instance, by implementing magnetic-field stabilization methods \cite{borkowski2023, cominotti2024, tiengo2025b} tailored to available bosonic mixture experiments \cite{lavoine2021, chisholm2022, cominotti2025}.

\vspace{1mm}
\textbf{Rabi-coupled bosonic mixtures.---}
We consider a bosonic mixture of two states, $\sigma = \uparrow,\downarrow$, driven by a Rabi coupling of frequency $\Omega$ and detuning $\delta$. 
The Hamiltonian, expressed in terms of the spinor field operator $\hat{\Psi}_{\mathbf{r}} = (\hat{\psi}_{\mathbf{r}\1}, \hat{\psi}_{\mathbf{r}\2})^T$, reads 
\begin{align}
\begin{split}
\hat{H}= \int  d\textbf{r} \, \bigg\{& \hat{\Psi}^\dagger_{\mathbf{r} } \, 
\left[ -\frac{\hbar^2\nabla^2_{\mathbf{r}}}{2m} + U_D(\mathbf{r}) + \hat{\xi} \right]  \hat{\Psi}_{ \mathbf{r}}
\\ 
&+ \sum_{ \sigma,\sigma' }
 \frac{g_{\sigma\sigma'}}{2}  \hat{\Psi}^\dagger_{\mathbf{r} \sigma} 
 \hat{\Psi}^\dagger_{\mathbf{r} \sigma'}   \hat{\Psi}_{\mathbf{r} \sigma} \hat{\Psi}_{\mathbf{r} \sigma'}   \bigg\},
 \label{generalH}
\end{split}
\end{align}
where the spin Hamiltonian $\hat{\xi} = -\hbar \, (\Omega \hat{\sigma}_x + \delta \hat{\sigma}_z)/2$ is expressed in terms of the Pauli matrices $\hat{\sigma}_i$, and the interaction strength is $g_{\sigma \sigma'} = 4 \pi \hbar^2 a_{\sigma\sigma'}/m$, with $m$ the atomic mass and $a_{\sigma\sigma'}$ the three-dimensional scattering lengths, and $\nabla_{\mathbf{r}}^2 = \nabla_{\boldsymbol{\rho}}^2 + \partial_z^2$, with $\boldsymbol{\rho} = (x,y)$. 
We treat within a unified framework the 3D, quasi-2D, and quasi-1D geometries by taking the external potential
\begin{equation}
U_D(\mathbf{r}) = \frac{m\omega_{\perp}^2}{2} \left[ (x^2+y^2) \, \delta_{1,D} + z^2 \, \delta_{2,D} \right],
\end{equation}
with $\delta$ the Kronecker delta, meaning that the system is harmonically confined with frequency $\omega_{\perp}$ for $D=1,2$ and unconfined for $D=3$.

The one-body part of the Hamiltonian \eqref{generalH} is diagonalized by a rotation in the spin basis. Denoting with $u = +,-$ the rotated spin states, we define the dressed spinor state $\hat{\Phi}_{\mathbf{r}} = (\hat{\phi}_{\vec{r}+}, \hat{\phi}_{\vec{r}-})^T$, related to the original spinor as \cite{sanz2022, chisholm2022, tiengo2025}
\begin{equation}
\hat{\Phi}_{\mathbf{r}} = R \, \hat{\Psi}_{\mathbf{r}}, \quad  R = \begin{pmatrix}
\sin\theta & \cos\theta
\\
-\cos\theta & \sin\theta
\end{pmatrix},
\end{equation}
where $\cos\theta = -\alpha_0/\sqrt{1+\alpha_0^2}$ and $\sin\theta = 1/\sqrt{1+\alpha_0^2}$ are expressed in terms of $\alpha_0 = \delta/\Omega + \sqrt{1+(\delta/\Omega)^2}$.
Note that the parameter $\alpha_0$ is related to the density imbalance minimizing the mean-field energy for weak interactions \cite{lavoine2021}. 
Substituting $\hat{\Psi}_{\mathbf{r}} = R^{T} \hat{\Phi}_{\mathbf{r}}$ in the Hamiltonian, we obtain 
\begin{align}
\begin{split}
\hat{H}= \int  d\textbf{r} \, \bigg\{& \hat{\Phi}^\dagger_{\mathbf{r} } \, 
\left[ -\frac{\hbar^2\nabla^2_{\mathbf{r}}}{2m} + U_D(\mathbf{r}) + \frac{\hbar{\Tilde{\Omega}}}{2} \hat{\sigma}_z \right]  \hat{\Phi}_{ \mathbf{r}}
\\ 
&+ { \sum_{ u_1,u_2,u_3,u_4 }
 \frac{g_{u_1 u_2}^{u_3 u_4}}{2}  \hat{\Phi}^\dagger_{\mathbf{r} u_1} 
 \hat{\Phi}^\dagger_{\mathbf{r} u_2}   \hat{\Phi}_{\mathbf{r} u_3} \hat{\Phi}_{\mathbf{r} u_4} }  \bigg\},
\end{split}
\end{align}
where ${\Tilde{\Omega}} = \sqrt{\Omega^2+\delta^2}$ is the energy gap between the dressed spin states $\ket{+} = \sin\theta \ket{\1} + \cos\theta \ket{\2}$ and $\ket{-} = - \cos\theta \ket{\1} + \sin\theta \ket{\2}$, and $g_{u_1 u_2}^{u_3 u_4}$ now represent the interaction strengths between atoms in the dressed spin states.

For small densities (or weak interactions), the Rabi-driven system forms a condensate of atoms in the state $\ket{-}$, interacting pairwise with strength $g_{--}^{--} = \cos^4\theta \, g_{\1\1} + \sin^4 \theta \, g_{\2\2} + \sin^2 2\theta \, g_{\1\2} /2$ \cite{lavoine2021, sanz2022}.
This result reproduces, in the uncoupled case $\Omega \to 0$, the familiar limits $g_{--}^{--} = g_{\1\1}$ for $\delta >0$ and $g_{--}^{--} = g_{\2\2}$ for $\delta <0$. 
This lowest-band approximation can well describe the equilibrium properties, yet the excited spin state $\ket{+}$ can play an important role in scattering problems \cite{sanz2022}, even at low energy.
In the rest of the paper, in particular, we analyze how higher spin channels affect the two-body scattering problem in $D=1,2,3$. 
For simplicity, we will work in the natural units $\hbar$, $m$ in $D=3$, and in $\hbar$, $m$, $l_{\perp}$ in $D = 1,2$, where $l_{\perp} = \sqrt{\hbar^2/(m\omega_{\perp})}$ is the harmonic-oscillator length.

\vspace{1mm}
\textbf{Exact solution of the two-body scattering problem.---}
The two-body scattering problem includes both spin and orbital degrees of freedom. 
Concerning the spin, we directly work with unnormalized dressed spin states, $\ket{-} = \ket{\1} + \ket{\2}/\alpha_0$ and $\ket{+} = \ket{\1} - \alpha_0 \ket{\2}$, and consider the three possible states for the two-bosons system \cite{lavoine2021},
\begin{align}
\begin{split}
\label{dressed2bodystates}
\ket{--} &= \ket{\uparrow\uparrow} + (\ket{\uparrow\downarrow} + \ket{\downarrow\uparrow})/\alpha_0+\ket{\downarrow\downarrow}/\alpha_0^2,
\\
\ket{+-} &= \ket{\uparrow\uparrow} -(\delta/\Omega) (\ket{\uparrow\downarrow} + \ket{\downarrow\uparrow})-\ket{\downarrow\downarrow},
\\
\ket{++} &= \ket{\uparrow\uparrow} - \alpha_0(\ket{\uparrow\downarrow} + \ket{\downarrow\uparrow})+\alpha_0^2\ket{\downarrow\downarrow},
\end{split}
\end{align}
where only symmetric spin states under exchange appear since the bosonic orbital wave function is symmetric \footnote{We use the state $\ket{+-}$ for simplicity of notation, here equivalent to the symmetrized one: $(\ket{+-} + \ket{-+})/\sqrt{2} \equiv \sqrt{2} \ket{+-}$. The antisymmetric state $(\ket{+-} - \ket{-+})/\sqrt{2}$ is instead incompatible with bosonic symmetry.}. 
Concerning the orbital motion, it is separable as usual into the free motion of the center of mass and into the relative scattering of a particle with reduced mass $\mu = 1/2$. 
Since the center-of-mass problem is trivial, we henceforth focus on the sole relative one.

We model the two-body scattering state as
\begin{align}
\begin{split}
\ket{\Phi(\mathbf{r})} = 
\phi_{--}(\mathbf{r}) \ket{--} 
+ \phi_{+-}(\mathbf{r}) \ket{+-} 
+ \phi_{++}(\mathbf{r}) \ket{++},
\label{Spinorscatteringstate}
\end{split}
\end{align}
where the orbital wave functions $\phi_{uu'}(\mathbf{r})$ solve the relative Schr\"odinger equation
\begin{equation}
\left[ - \frac{\nabla_{\mathbf{r}}^2}{2\mu} + u_D(\mathbf{r})  - E_{uu'} \right] \phi_{uu'}(\mathbf{r})=0,
\label{Schrsinglecomp}
\end{equation}
with $u_D(\mathbf{r}) = \mu \left[ (x^2+y^2) \, \delta_{1,D} + z^2 \, \delta_{2,D} \right]/2$ and boundary conditions specified below. 
Note that the $s$-wave solutions of Eq.~\eqref{Schrsinglecomp} have already been obtained for a single-component gas in the context of scattering theory in 3D \cite{Landau1991}, quasi-2D \cite{petrovshlyapnikov}, and quasi-1D \cite{olshanii1998, bergeman2003}.
We now adapt them to solve the scattering problem of the quasi-low-dimensional two-components mixture {(see Ref.~\cite{petrov2014} for solutions in the purely $D$-dimensional case)}.

\begin{figure}[hbtp]
\centering
\includegraphics[width=0.98\columnwidth]{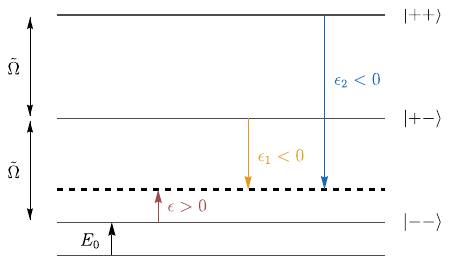}
\caption{Energy scheme representing the scattering process modeled in this paper. 
Physically, two $\ket{-}$ atoms scatter at the relative collisional energy $\epsilon$ and, as a result, can be virtually excited to the spin states $\ket{+-}$ or $\ket{++}$, in which their energies $\epsilon_1$ and $\epsilon_2$ are negative, and therefore correspond to bound states.
If the atoms are confined in quasi-low-dimensional geometries, the energies include the zero-point oscillator energy $E_0$, defined in the main text. 
We measure energies with respect to the one of two nontrapped atoms with zero relative kinetic energy.
}
\label{fig1}
\end{figure}

As illustrated in Fig.~\ref{fig1}, the scattering process involves the open channel $\ket{--}$ and the closed channels $\ket{+-}$ and $\ket{++}$.
The incoming wave in the open channel $\ket{--}$ has wave function $\phi_{--}(\mathbf{r})$ and energy $E_{--} = E_{0} + \epsilon$, with $\epsilon = q^2/(2\mu)$ being the positive collisional energy and $q$ the relative momentum.
Here $\epsilon < {\Tilde{\Omega}}$ and $E_{0} = (3-D)/2$ is the zero-point energy in quasi-$D$ dimensions.
The wave functions and energies of the closed channels are, respectively, $\phi_{+-}(\mathbf{r})$, $\phi_{++}(\mathbf{r})$, and $E_{+-} = E_{0} +\epsilon_1$, $E_{++} = E_{0} +\epsilon_2$, where $\epsilon_{1} = \epsilon - {\Tilde{\Omega}}$ and $\epsilon_{2} = \epsilon - 2{\Tilde{\Omega}}$ are negative, and correspond to the binding energies of the molecular states.

The open channel wave function is a superposition of a (propagating) regular solution plus a (scattered) singular solution, while the closed channels only include the singular component:
\begin{align}
\begin{split}
\phi_{--}(\mathbf{r}) &= \phi_{\rm reg}(\mathbf{r}) + C_{--} \, K_{E_{--}}(\mathbf{r}),
\\
\phi_{+-}(\mathbf{r}) &= C_{+-} \, K_{E_{+-}}(\mathbf{r}),
\\
\phi_{++}(\mathbf{r}) &= C_{++} \, K_{E_{++}}(\mathbf{r}),
\label{spinorOrbitalwfs}
\end{split}
\end{align}
where the $C_{uu'}$ are coefficients. 
In particular, the incoming regular $s$-wave solution reads
\begin{equation}
\phi_{\rm reg}(\mathbf{r}) = \begin{cases}
      \cos(q z) \, e^{-\rho^2/4}, \ & D=1,
      \\
      J_0(q\rho) \, e^{-z^2/4}, \ & D=2,
      \\
      \sin(qr)/(qr),  \ & D=3,
    \end{cases}
    \label{psireg}
\end{equation}
with $J_0(q\rho)$ the Bessel's function, while the scattered singular solution is
\begin{equation}
K_{E}(\mathbf{r}) = \begin{cases}
      \frac{1}{(4\pi)^{1/2}}
\int_0^{\infty} \frac{d\tau}{\sqrt{\tau}\sinh\tau} e^{-\frac{z^2}{4\tau}-\frac{\rho^2}{4\tanh\tau}+E\tau}, \ & D=1
      \\
      \frac{1}{(4\pi)^{1/2}}
\int_0^{\infty} \frac{d\tau}{\tau\sqrt{\sinh\tau}} e^{-\frac{\rho^2}{4\tau} -\frac{z^2}{4\tanh\tau}+E\tau}, \ & D=2
      \\
      \frac{e^{-\sqrt{-E} r}}{r},  \ & D=3  
    \end{cases}
    \label{KallDimensions}
\end{equation}
and the $D=1,2$ forms are obtained with the help of the harmonic-oscillator Green's function \cite{feynmanhibbs} (see also Appendix \ref{appB}). 
We emphasize that, while the regular part of the $D=1,2$ solution is in the transverse ground state, the scattered waves $K_{E}(\mathbf{r})$ include all transversely excited components.
As usual in scattering theory, we now determine the coefficients $C_{uu'}$ by expanding the state of Eq.~\eqref{Spinorscatteringstate}, with components given by Eqs.~(\ref{spinorOrbitalwfs}-\ref{KallDimensions}), at small and large interparticle separation, imposing that it matches the expected asymptotic forms. 

\vspace{0mm}
\textit{Short-distance expansion.---}
At short interparticle distance, we can simply replace Eq.~\eqref{psireg} with $\phi_{\rm reg}(0) = 1$, while Eq.~\eqref{KallDimensions} behaves as 
\begin{equation}
K_{E}(\mathbf{r}) = \frac{1}{r} + h_D(E-E_0),
\end{equation}
where the function $h_D(x)$ is obtained from the known solutions of the (quasi)-$D$-dimensional single-species problems (see Refs.~\cite{olshanii1998, petrovshlyapnikov, pricoupenko2007b} for additional details). 
In particular, $h_1(x) = \zeta(1/2,-x/2)/\sqrt{2}$, with $\zeta$ the Hurwitz zeta function, while \footnote{The formula for $h_2(x)$ converges slowly, but it can be computed more efficiently via analytical continuation \cite{pricoupenko2007b}.}
\begin{equation}
h_2(x)=\lim_{N\to\infty}\!\Bigl[\tfrac{\sqrt{2N}}{\pi}\ln\tfrac{N}{e^2}
-\sum_{j=0}^N \tfrac{(2j-1)!!}{(2j)!!} \tfrac{\ln(j-x/2-i0^+)}{\sqrt{2\pi}}\Bigr],
\label{wfunction}
\end{equation}
and $h_3(x) = - \sqrt{-x}$, where we choose the convention $\sqrt{-|x|} = -i \sqrt{|x|}$.
These expansions provide the behavior of the state \eqref{Spinorscatteringstate} at short distance. 
Imposing the $s$-wave Bethe-Peierls boundary conditions for scattering in the bare spin states 
\begin{equation}
\braket{\sigma \sigma'}{\Phi(\mathbf{r})} \propto \frac{1}{r} - \frac{1}{a_{\sigma\sigma'}}, %+ R_{\sigma\sigma'}^* \epsilon,
\end{equation}
we obtain a system of algebraic equations for the coefficients $C_{--}$, $C_{+-}$, and $C_{++}$:
\begin{widetext}
\begin{equation}\label{matrix}
\begin{pmatrix}
a_{\1\1}^{-1} + h_D(\epsilon)
& a_{\1\1}^{-1} + h_D(\epsilon_1)
& a_{\1\1}^{-1} + h_D(\epsilon_2)
\\
a_{\1\2}^{-1} + h_D(\epsilon)
& -\alpha_0(\delta/\Omega)[a_{\1\2}^{-1}+ h_D(\epsilon_1) ] 
& -\alpha_0^2 [a_{\1\2}^{-1} + h_D(\epsilon_2)]
\\
a_{\2\2}^{-1} + h_D(\epsilon) 
& -\alpha_0^2[a_{\2\2}^{-1}+ h_D(\epsilon_1) ]
& \alpha_0^4 [a_{\2\2}^{-1} + h_D(\epsilon_2)]
\end{pmatrix}
\begin{pmatrix}
C_{--}\\C_{+-}\\C_{++}
\end{pmatrix}
= -
\begin{pmatrix}
1\\1\\1
\end{pmatrix}.
\end{equation}
\end{widetext}
This matrix equation can be easily solved to get the analytical expressions of the $C_{uu'}$ coefficients. 
Actually, we will only need $C_{--}$ to evaluate the $s$-wave scattering amplitudes, obtained below by expanding the scattering state at large distance.

\textit{Large-distance expansion.---}
At large distance, the $\ket{--}$ component of the two-body scattering state \eqref{Spinorscatteringstate} has an oscillating behavior, while the $\ket{+-}$ and $\ket{++}$ components vanish.
Therefore, the wave function tends asymptotically to $\phi_{--}(\mathbf{r})$, and its projection on the transverse ground-state wave function is proportional to
\begin{equation}
\begin{cases}
      \cos(q z) + \frac{iC_{--} }{q} \, e^{iq |z| }, \ & D=1,
      \\
      J_0(q\rho) + i C_{--} \sqrt{\frac{\pi}{2}} \, H_0^{(1)}(q\rho), \ & D=2,
      \\
      \frac{\sin(qr)}{qr} + C_{--} \frac{e^{iqr}}{r}, \ & D=3,
    \end{cases}
\end{equation}
where $H_0^{(1)}(q\rho)$ is the Hankel function.
By comparing it with the asymptotic $s$-wave solutions of the two-body Schr\"odinger equations [cf. Eqs.~\eqref{psi1D}, \eqref{psi2D}, and \eqref{psi3D}],
we find the $s$-wave scattering amplitudes
\begin{equation}
f_0(q) = 
\begin{cases}
      iC_{--}/q, \qquad &D=1,
      \\
      - \sqrt{8\pi} \, C_{--}, \qquad &D=2,
      \\
      C_{--}, \qquad &D=3.
    \end{cases}
\end{equation}
Finally, by comparing the scattering amplitudes with the known $D$-dimensional expressions [cf. Eqs. \eqref{f1D}, \eqref{f2D}, and \eqref{f3D}], we can express the $D$-dimensional scattering lengths directly in terms of $C_{--}$ as
\begin{align}
\begin{split}
a_{1D} &= \frac{1}{C_{--}} + \frac{i}{q},
\\
a_{2D} &= \frac{2}{q}\exp(\sqrt{\frac{\pi}{2}}\frac{1}{C_{--}} + \frac{i\pi}{2}-\gamma_{E}),
\\
a_{3D} &= - \left( \frac{1}{C_{--}} + i q \right)^{-1}.
\label{aDdimensions}
\end{split}
\end{align}

\begin{figure}[hbtp]
\centering
\includegraphics[width=0.98\columnwidth]{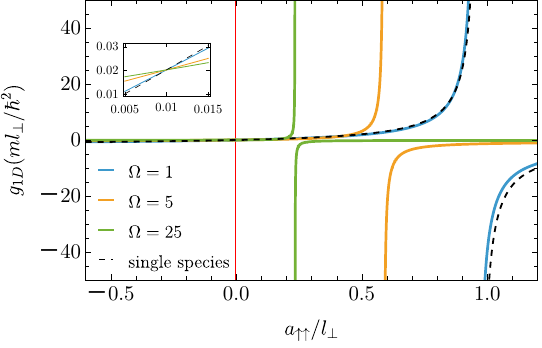}
\caption{Effective one-dimensional interaction strength versus the scattering length $a_{\1\1}$, obtained by evaluating Eq.~\eqref{g1Dmixture} for parameters similar to those of Ref.~\cite{lavoine2021}: $\epsilon = 0^{+}$,  $a_{\1\2} = a_{\2\2} = 0.01$, and $\delta = 2$.
The dashed line represents the interaction strength in a quasi-1D single-species calculated in Ref.~\cite{olshanii1998}, i.e., Eq.~\eqref{g1DOlshanii} for $\sigma = \1$, and it is reproduced {in the weak driving limit $\Omega/\delta \to 0$}.
Note that one can conveniently shift the resonance position towards smaller scattering lengths by increasing $\Omega$, reaching for $\Omega \gg |\delta|$ the red vertical line [corresponding to $1/\bar{a}_+=-\zeta(1/2,1)/\sqrt{2}$]. 
Inset: magnification around the point where all scattering lengths are equal, where Eq.~\eqref{g1DOlshanii} is exact irrespective of $\Omega$.
}
\label{fig2}
\end{figure}

\vspace{1mm}
\textbf{Results.---}
In the quasi-1D geometry the effective one-dimensional scattering strength is
\begin{equation}
g_{\text{1D}} = -\frac{1}{\mu a_{\text{1D}}},
\label{g1Dmixture}
\end{equation}
where $a_{\text{1D}}$ is defined in Eq.~\eqref{aDdimensions}.
The resulting expression for $g_{\text{1D}}$ is bulky but nonetheless analytical, and depends on the problem parameters $\epsilon$, $a_{\sigma\sigma'}$, $\delta$, $\Omega$, and $l_{\perp}$. 
We plot $g_{\text{1D}}$ versus $a_{\1\1} / l_{\perp}$ for different values of $\Omega$ in Fig.~\ref{fig2}.
The solid curves show confinement-induced resonances which can be displaced to different locations depending on the values of $\delta$ and $\Omega$. To understand their behavior, we now analyze the different limits.

Let us first recall Olshanii's result \cite{olshanii1998} for scattering in a single-species gas:
\begin{equation}
g_{O}(a) =\frac{2}{a^{-1} +  \zeta(1/2,1-\epsilon/2)/\sqrt{2}}.
\label{g1DOlshanii}
\end{equation}
In the weak driving (or equivalently off-resonant) limit $\Omega/\delta \to 0$,
the normalized scattering state $\ket{--}$ tends to $\ket{\sigma\sigma}$, with $\sigma = \1$ for $\delta >0$ and $\sigma = \2$ for $\delta <0$. 
The interaction strength reduces to {$g_{\text{1D}} \xrightarrow[]{\Omega/\delta \to 0}  g_{O}(a_{\sigma\sigma})$} for $0 < \epsilon < 1$ \footnote{Note that we used here the series definition of the Hurwitz zeta function to express $\zeta(1/2,-\epsilon/2)/\sqrt{2} = i/\sqrt{\epsilon} + \zeta(1/2,1-\epsilon/2)/\sqrt{2}$ for $0 < \epsilon < 1$}. 
In particular, for $\epsilon = 0^{+}$ one gets the conventional confinement-induced resonance around the scattering length $a_{\sigma\sigma}/l_{\perp} \sim 0.968$.
This behavior is clearly shown in Fig.~\ref{fig2}, where for $\Omega \ll |\delta|$ the interaction strength converges to the single-species result of Eq.~\eqref{g1DOlshanii} (dashed line).

In the limit of large ${\Tilde{\Omega}}$ {at fixed $\delta/\Omega$ (i.e., at fixed $\theta$)}, we obtain
\begin{align}
\begin{split}
g_{\text{1D}} & \!\xrightarrow[]{{\Tilde{\Omega}} \to \infty}  g_O(\bar{a}^+_{\theta})-  \frac{g_O(\bar{a}^+_{\theta})^2}{4\sqrt{{\Tilde{\Omega}}}} \! \{\sqrt{2} \sin^4(2\theta)(\Bar{a}^{-}_{\pi/4})^2  + \\& 
\sin^2(2\theta)[\cos\theta^2(a_{\1\1}^{-1}-a_{\1\2}^{-1}) + \sin^2\theta(a_{\1\2}^{-1}-a_{\2\2}^{-1})]^2  \},
\label{g1DlargeTildeOmega}
\end{split}
\end{align}
with $\Bar{a}^{\pm}_{\theta} = [a_{\1\1}^{-1} \cos^4\theta \pm a_{\1\2}^{-1} \sin^2(2\theta)/2 + a_{\2\2}^{-1} \sin^4\theta]^{-1}$.
The two addends in the curly brackets correspond to virtual second-order processes where the colliding $\ket{--}$ atoms respectively populate the $\ket{++}$ and $\ket{+-}$ spin states, in close analogy to what is discussed in Ref.~\cite{tiengo2025}.
In the strong-driving limit $\Omega \gg |\delta|$ the spin states become $\ket{\pm} = (\ket{\1} \mp \ket{\2})/\sqrt{2}$, and the interaction strength reads
\begin{align}
\begin{split}
g_{\text{1D}} \!\xrightarrow[]{\Omega \to \infty} \! g_O(\bar{a}^+_{\pi/4})-  \frac{g_O(\bar{a}^+_{\pi/4})^2}{\sqrt{8\Omega}} \! \left[(\Bar{a}^{-}_{\pi/4})^2 \! + \! \frac{(a_{\1\1}^{-1}-a_{\2\2}^{-1})^2}{4\sqrt{2}}\right].
\label{g1DlargeOmega}
\end{split}
\end{align}
By keeping $a_{\1\2}$ and $a_{\2\2}$ fixed, and varying $a_{\1\1}$, the $\epsilon = 0^{+}$ resonance is centered around $a_{\1\1} = -[a_{\2\2}^{-1}+2 a_{\1\2}^{-1} + 2 \sqrt{2} \, \zeta(1/2,1)]^{-1}$ for infinite $\Omega$.
Note that this value, depicted as the red vertical line in Fig.~\ref{fig2}, can take very small values for small $a_{\1\2}$ and $a_{\2\2}$. 
Furthermore, numerical explorations for realistic parameters \cite{derrico2007} confirm that fine-tuned combinations of small scattering lengths $a_{\sigma\sigma'} \ll l_{\perp}$ can produce resonances when $\Omega \gg 1$.
{For} moderate $\Omega$ values, the resonance is still displaced towards $a_{\1\1} < l_{\perp}$, as Fig.~\ref{fig2} shows.
We also mention that, for resonant couplings $|\delta/\Omega| \lesssim 1$, tuning $\delta$ changes appreciably the spin composition and therefore can further contribute to shift the resonance location quantitatively.
These effects can facilitate the observation of confinement-induced resonances in experimental setups \cite{lavoine2021, sanz2022}, and it can allow a better control of the scattering properties in ultracold bosonic mixtures.

\begin{figure}[hbtp]
\centering
\includegraphics[width=0.98\columnwidth]{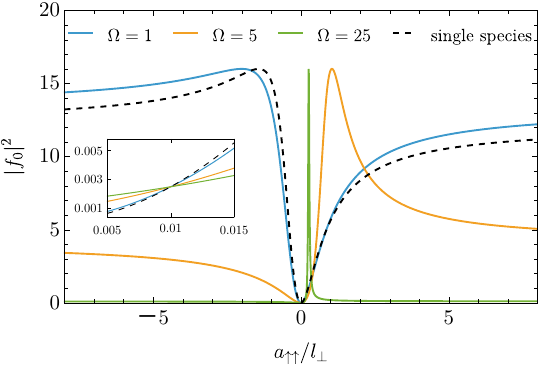}
\caption{Two-dimensional s-wave scattering amplitude versus the scattering length, evaluated at the scattering energy $\epsilon = 0.05$ and setting the other parameters as in Fig.~\ref{fig2}.
The dashed line is Eq.~\eqref{Petrovf0}, obtained for a single-species gas, and it is reproduced for small Rabi frequency {$\Omega/\delta \to 0$}.
Inset: plot magnification around the region where all scattering lengths are equal, where Eq.~\eqref{Petrovf0} holds independent of $\Omega$. 
}
\label{fig3}
\end{figure}

In the quasi-2D geometry, we directly evaluate the $s$-wave scattering amplitude $f_0 = - \sqrt{8\pi} \, C_{--}$.
We plot $|f_0|^2$ versus $a_{\1\1}/l_{\perp}$ for different choices of $\Omega$ in Fig.~\ref{fig3}.
Similar to the quasi-1D case, at weak driving {$\Omega/\delta \to 0$} the scattering amplitude yields Petrov and Shlyapnikov's result  \cite{petrovshlyapnikov} 
\begin{equation}
|f_0(q^2)|^2 = \frac{16 \pi^2}{\left[\sqrt{2\pi}/a_{\sigma\sigma} + 
\ln \left(B/\pi q^2\right)\right]^2 + \pi^2},
\label{Petrovf0}
\end{equation}
where $B= 0.9049$ \cite{pricoupenko2007, pricoupenko2007b}, and again $\sigma = \1$ for $\delta >0$, while $\sigma = \2$ for $\delta <0$.
At strong driving, $\Omega \gg |\delta|$, we then observe a displacement of the scattering resonance towards small values of $a_{\sigma\sigma'}/l_{\perp}$. Since the typical experimental scenario corresponds to $a_{\1\1} \ll l_{\perp}$, our results enable probing and exploiting more easily the confinement-induced resonances in ultracold quantum gases.

Concerning the 3D results, we obtain the 3D coupling constant $g_{3D} = 4 \pi a_{3D}$ and we reproduce the findings of Ref.~\cite{lavoine2021}, to which we refer for detailed analyses. 

Finally, let us discuss what occurs when all scattering lengths are equal, i.e., $a_{\sigma \sigma'} = a$.
In this case, interactions become spin independent as all scattering channels are characterized by the same Bethe-Peierls boundary conditions.
Any mixing of the spin states, realized by varying the Rabi coupling $\Omega$, does not affect the scattering problem, which is exactly described by the single-species results of Eqs.~\eqref{g1DOlshanii} and \eqref{Petrovf0}.
We demonstrate this result in Figs.~\ref{fig2} and \ref{fig3} insets, which magnify the region where all scattering lengths coincide.

\vspace{1mm}
\textbf{Conclusions.---}
We analyzed the scattering problem in low-dimensional bosonic mixtures with Rabi coupling, finding that the driving field can displace the resonance location towards scattering lengths much smaller than the oscillator length.
In particular, in the large Rabi coupling regime we observe that confinement-induced resonances can be engineered for fine-tuned small scattering lengths $a_{\sigma\sigma'} \ll l_{\perp}$.
Our findings open the way towards the observation of confinement-induced resonances in ultracold atomic mixtures
and provide a handle to efficiently control interactions in atomic gases \cite{lamporesi2010, zurn2012}.
We presented our results choosing realistic values for a recent experiment~\cite{lavoine2021} and latest improvements in the stability of the driving field \cite{borkowski2023, cominotti2024, tiengo2025b} combined with strong transverse confinement could further facilitate their observation.

\vspace{-5mm}
\begin{acknowledgements}
We gratefully thank T. Bourdel, J. Levinsen, M. Parish, D.~S.~Petrov and L. Tarruell for inspiring discussions, and acknowledge contributions of D.~S.~Petrov to the solution of the quasi-1D problem.
A.~T. acknowledges funding by the European Union under the Horizon Europe MSCA programme via the project 101146753 (QUANTIFLAC).
P.~M. acknowledges support by the ICREA {\it Academia} program.
We further acknowledge financial support by the Spanish Ministerio de Ciencia, Innovación y Universidades (grant PID2023-147469NB-C21, financed by MICIU/AEI/10.13039/501100011033 and FEDER-EU).
\end{acknowledgements}

\appendix

\section{Single-species $D$-dimensional $s$-wave scattering problems} 

Here we review the $D$-dimensional problems of two identical bosons scattering at positive energy $\epsilon=q^2/(2\mu)$.

\textbf{1D case.---} The 1D Schr\"odinger equation in rescaled relative coordinates reads
\begin{equation}
\left[ - \frac{\partial_z^2}{2\mu} + g_{\text{1D}} \delta(z)  - \epsilon \right] \psi(z)=0,
\label{eqschr2}
\end{equation}
where $g_{\text{1D}}$ is the 1D interaction strength.
Equivalently, we solve $(-\partial_z^2 - q^2)\psi(z)=0$ equipped with the relation $\partial_z \psi(z) |^{z=0^+}_{z=0^-} = 2\mu g_{\text{1D}} \, \psi(0)$.
The solution of even symmetry under $z \to -z$ reflection can be expressed as
\begin{align}
\begin{split}
\psi(z) 
&\propto \cos(q z) + \frac{\mu g_{\text{1D}}}{q} \sin(q |z|)
\\
&\propto 
\cos(q z) + f_{1D} (q) \, e^{i q |z|},
\label{psi1D}
\end{split}
\end{align}
where we introduce the 1D scattering amplitude $f_{1D}$, related to the 1D interaction strength as $g_{\text{1D}} = - (q/\mu) \Re(f_{1D})/\Im(f_{1D})$ \cite{bergeman2003}.
If we require the solution to satisfy the Bethe-Peierls boundary condition $\psi(z \to 0) \propto 1 - |z|/a_{\text{1D}}$, with $a_{\text{1D}}$ the one-dimensional $s$-wave scattering length, we find $g_{\text{1D}} = -1/(\mu a_{\text{1D}})$, and obtain 
\begin{equation}
f_{1D} (q) = - \frac{1}{1+i q a_{\text{1D}}}.
\label{f1D}
\end{equation}

\textbf{2D case.---} The 2D $s$-wave Schr\"odinger equation in rescaled relative coordinates reads 
\begin{equation}
\left[ - \frac{\partial_\rho^2 + \rho^{-1}\partial_\rho}{2\mu}  - \epsilon \right] \psi(\rho)=0.
\end{equation}
The solution is given by \cite{Landau1991, petrovshlyapnikov}
\begin{align}
\begin{split}
\psi(\rho) = J_0(q\rho) - i \frac{f_{2D}(q)}{4} H_0^{(1)}(q\rho),
\label{psi2D}
\end{split}
\end{align}
and imposing the Bethe-Peierls boundary condition $\psi(\rho \to 0) \propto \ln(\rho/a_{2D})$, with $a_{2D}$ the two-dimensional $s$-wave scattering length, we obtain the  $s$-wave scattering amplitude
\begin{equation}
f_{2D} (q) = \frac{2\pi}{-\ln(q a_{2D} e^{\gamma_E}/2)+i\pi/2},
\label{f2D}
\end{equation}
with $\gamma_E \approx 0.577$ the Euler-Mascheroni constant. 

\textbf{3D case.---} 
The 3D Schr\"odinger equation for the relative coordinate is
\begin{equation}
\left(-\frac{\nabla_{\mathbf{r}}^2}{2\mu} - \epsilon \right)\psi({\mathbf{r}}) = 0,
\label{quasi2Dschrodinger}
\end{equation}
and admits the asymptotic solution $\psi(\mathbf{r}) \propto e^{iq z} + f_{3D}(q,\theta) \, e^{iqr}/r$, with $f_{3D}(q,\theta)$ the scattering amplitude including all partial waves \cite{Landau1991}.
We are interested in the $s$-wave asymptotic component, which we calculate by integrating the previous solution over all solid angles, obtaining
\begin{equation}
\psi(r) \propto \frac{\sin(qr)}{qr} + f_{3D}(q) \frac{e^{iqr}}{r}.
\label{psi3D}
\end{equation}
By imposing the $s$-wave Bethe-Peierls boundary condition $\psi(r \to 0)\propto 1/r - 1/a_{3D}$, we obtain 
\begin{equation}
f_{3D}(q) = - \frac{1}{1/a_{3D}+iq}.
\label{f3D}
\end{equation}

\section{Green's functions}
\label{appB}
Given the Schr\"odinger equation $(\hat{H}_{\mathbf{r}} - E) \psi(\mathbf{r}) = 0$ with time-independent Hamiltonian $\hat{H}_{\mathbf{r}}$, we define the Green's function in the energy space $K_E(\mathbf{r},\mathbf{r}')$ as the function solving $(\hat{H}_{\mathbf{r}} - E) K_E(\mathbf{r},\mathbf{r}') = \delta(\mathbf{r}-\mathbf{r}')$. To evaluate it explicitly, we express it as \cite{feynmanhibbs}
\begin{equation}
K_E(\mathbf{r},\mathbf{r}') = \int_{-\infty}^{\infty} d\tau \, e^{E\tau} K(\mathbf{r},\mathbf{r}',\tau),
\label{KE}
\end{equation}
where $K(\mathbf{r},\mathbf{r}',\tau)$ is the imaginary-time propagator bringing the system from $\mathbf{r}'$ to $\mathbf{r}$ in the time $\tau$. 
This function reads \cite{feynmanhibbs}
\begin{equation}
K(\mathbf{r},\mathbf{r}',\tau) = \sum_n \phi_n(\mathbf{r}) \phi_n^{*}(\mathbf{r}') e^{-E_n \tau}
\end{equation}
for $\tau > 0$, and $K(\mathbf{r},\mathbf{r}',\tau) = 0$ for $\tau <0$, where $\hat{H}_{\mathbf{r}} \phi_n(\mathbf{r}) = E_n \phi_n(\mathbf{r})$.

A useful property involves Hamiltonians that sum commuting operators, for instance in the form $\hat{H}_{\mathbf{r}} = \hat{H}_{\rho} + \hat{H}_{z}$ that appears in this paper. In this case, the imaginary-time propagator factorizes as $K(\rho,z,\rho',z',\tau) = K^{(\rho)}(\rho,\rho',\tau) K^{(z)}(z,z',\tau)$.
Thanks to this property, in the main text we obtain the Green's functions of the type \eqref{KE} by multiplying the known forms of the imaginary-time propagators for the free particle and for the harmonic oscillator with reduced mass.
In particular, for initial positions $\mathbf{r}'=0$, the $\tau>0$ expressions for the free 1D, and harmonically trapped 1D and 2D cases are \cite{feynmanhibbs}:
\begin{align}
&\hat{H}_z=-\partial_z^2 
&\rightarrow\quad
& K^{(z)}(z,\tau) = \frac{e^{-\frac{z^2}{4\tau}}}{\sqrt{4\pi\tau}}
\\
&\hat{H}_x=-\partial_x^2+\frac{x^2}{4} 
&\rightarrow\quad
&  K^{(x)}(x,\tau) = \frac{e^{-\frac{x^2}{4\tanh{\tau}}}}{\sqrt{4\pi\sinh{\tau}} }
\nonumber
\\
&\hat{H}_{\rho}=-\partial_{\rho}^2-\frac{\partial{\rho}}{\rho}+\frac{\rho^2}{4} 
&\rightarrow\quad
& K^{(\rho)}(\rho,\tau) = \frac{e^{-\frac{\rho^2}{4\tanh{\tau}}}}{4\pi\sinh{\tau}}.
\nonumber
\end{align}

\end{document}